\documentclass[conference]{IEEEtran}
\IEEEoverridecommandlockouts
\usepackage{cite}
\usepackage{amsmath,amssymb,amsfonts}
\usepackage{algorithmic}
\usepackage{graphicx}
\usepackage{textcomp}
\usepackage{multirow}
\usepackage{booktabs}
\usepackage{hyperref}
\usepackage{xcolor}
\def\BibTeX{{\rm B\kern-.05em{\sc i\kern-.025em b}\kern-.08em
    T\kern-.1667em\lower.7ex\hbox{E}\kern-.125emX}}
\begin{document}
%
%
\setlength{\lightrulewidth}{0.02em}
\setlength{\heavyrulewidth}{0.08em}
%
\setlength{\textfloatsep}{2.0pt plus 3.0pt minus 8.0pt}
\setlength{\floatsep}{2.0pt plus 3.0pt minus 8.0pt}
\setlength{\intextsep}{3.0pt plus 3.0pt minus 8.0pt}
\setlength{\dbltextfloatsep}{4.0pt plus 3.0pt minus 8.0pt}
%
%
\title{Generative Speech Foundation Model Pretraining for High-Quality Speech Extraction and Restoration
}
%
\author{
    \IEEEauthorblockN{
        Pin-Jui Ku\IEEEauthorrefmark{1}\IEEEauthorrefmark{2},
        Alexander H. Liu\IEEEauthorrefmark{1}\IEEEauthorrefmark{3}, 
        Roman Korostik\IEEEauthorrefmark{1}, 
        Sung-Feng Huang\IEEEauthorrefmark{1}, 
        Szu-Wei Fu\IEEEauthorrefmark{1} and 
        Ante Juki\'{c}\IEEEauthorrefmark{1}
    }
    \IEEEauthorblockA{
        \IEEEauthorrefmark{1}NVIDIA, USA \quad \IEEEauthorrefmark{2}Georgia Institute of Technology, USA \quad \IEEEauthorrefmark{3}Massachusetts Institute of Technology, USA\\
    }
}

\maketitle
\begin{abstract}
This paper proposes a generative pretraining foundation model for high-quality speech restoration tasks. By directly operating on complex-valued short-time Fourier transform coefficients, our model does not rely on any vocoders for time-domain signal reconstruction. As a result, our model simplifies the synthesis process and removes the quality upper-bound introduced by any mel-spectrogram vocoder compared to prior work SpeechFlow. The proposed method is evaluated on multiple speech restoration tasks, including speech denoising, bandwidth extension, codec artifact removal, and target speaker extraction. In all scenarios, finetuning our pretrained model results in superior performance over strong baselines. Notably, in the target speaker extraction task, our model outperforms existing systems, including those leveraging SSL-pretrained encoders like WavLM. The code and the pretrained checkpoints are publicly available in the NVIDIA NeMo framework.

\end{abstract}


\section{Introduction}
\label{sec:intro}
Speech restoration (SR) aims at improving the quality and intelligibility of a corrupted speech observation by removing the undesired components, such as background noise and reverberation of the recording environment, or interfering speech signals.
Furthermore, SR also aims to reduce other corruptions, such as audio compression, coding artifacts, and bandwidth limitations, which can cause information loss and further degrade the quality of the speech signal.

In recent years, deep neural network (DNN) based approaches have demonstrated superiority over conventional signal processing techniques on many SR tasks.
They can be broadly categorized into two categories: deterministic models and generative models. 
On the one hand, deterministic models aim to provide an estimate of the clean signal from the corrupted signal, e.g., using a deep model to estimate real or complex valued spectral components~\cite{xu2014regression, han2015learning}, masks~\cite{wang2008time,luo2019conv,ku23s4nd}, or the time domain signal~\cite{Dfossez2020}.
On the other hand, generative models aim to learn a prior distribution over the clean signal, e.g., using variational autoencoders~\cite{Bando2020adaptive, Fang2021variational}, generative adversarial networks~\cite{pascual2017segan}, and diffusion models~\cite{Lu2022conditional, yen2022cold,lemercier2023analysing,lemercier2023storm}.
Although generative SR models have gained more popularity in recent years, modern speech enhancement models with state-of-the-art performance remain deterministic-based with a distinctive performance gap between deterministic-based and generative-based models~\cite{chao2024investigation, lu2023mp}.

A pretraining framework for generative models using flow matching, named SpeechFlow, has been proposed in~\cite{liu2024generative}.
While SpeechFlow demonstrated great potential in improving speech generation abilities~\cite{le2023voicebox, vyas2023audiobox}, it has several drawbacks. Firstly, SpeechFlow operates in the \mbox{mel-spectrogram} domain and requires an additional vocoder to convert the generated \mbox{mel-spectrogram} signals into the corresponding time domain signals.
Therefore, its performance is limited by an upper bound introduced by the vocoder. Furthermore, different vocoders had to be employed according to the different downstream tasks. This complicates the overall generation process and makes it challenging to adapt SpeechFlow to different speech extraction and restoration tasks.  

In this paper, we propose improvements to pretrain a foundation generative model and demonstrate its effectiveness for speech processing.
First, we perform flow matching directly with complex short-time Fourier transform (STFT) coefficients, eliminating the need for task-specific vocoders to generate time-domain signals. This greatly simplifies the generation process and improves speech quality. 
Unlike~\cite{liu2024generative}, the proposed model can be used without modifications in different speech processing tasks.
Next, we finetune the pretrained model on various speech restoration tasks, including speech denoising, codec artifact removal, bandwidth extension, and target speaker extraction (TSE). 
The finetuned model is evaluated against strong baseline models for each task.
On the speech denoising task, the proposed model outperforms SpeechFlow and existing generative-based speech enhancement systems in terms of speech quality metrics, narrowing the performance gap between deterministic and generative-based speech denoising models.
For codec artifact removal, the proposed model restores high-quality speech from low-bitrate codec output using only four codebooks.
The generated speech is of comparable quality to that of a three-times higher bitrate codec using 12 codebooks.
On the TSE task, the proposed model significantly outperforms existing TSE systems on the Libri2Mix dataset.
Finally, the implementation and a pretrained checkpoint of the proposed model are open-sourced in NVIDIA NeMo framework~\cite{nemo_toolkit}. 

%

%
\section{Models}
\label{sec:method}
\subsection{Flow matching generative model pretraining}
The flow matching~\cite{lipman2023flow} generative model belongs to the family of continuous normalizing flows~\cite{chen2018Neural}.
It defines a time-dependent probability density path $p_t: [0, 1] \times \mathbb{R}^d \rightarrow \mathbb{R}_{0+}$ in the data space $\mathbb{R}^d$ from a simple starting prior $p_0$ to the final target distribution $p_1$, and a time-dependent vector field $v_t: [0, 1] \times \mathbb{R}^d \rightarrow \mathbb{R}^d$.
The vector field $v_t$ is used to construct a time-dependent diffeomorphic map, called a flow, $\phi_t: [0,1]\times \mathbb{R}^d \rightarrow \mathbb{R}^d$  defined by the following ordinary differential equation (ODE):
\begin{gather}
    \label{eq:ode}
    \frac{d}{dt}\phi_t \left( \mathbf{x} \right) = v_t \left( \phi_t \left( \mathbf{x} \right) \right) , \quad \phi_0(\mathbf{x}) = \mathbf{x} ,
\end{gather}
with $\mathbf{x} \in \mathbb{R}^d$.
Furthermore, a vector field $v_t$ is said to generate a probability path $p_t$ if $p_t$ and the flow $\phi_t$ constructed by $v_t$ satisfy the following equation:
\begin{equation}
        p_t \left( \mathbf{x} \right) = p_0 \left( \phi_t^{-1} \left( \mathbf{x} \right) \right) \text{det} \left( \frac{\partial \phi_t^{-1}}{\partial \mathbf{x}} ( \mathbf{x} ) \right)
\end{equation}
Now, given the target vector field $v_t(\mathbf{x})$ that generates $p_t(\mathbf{x})$, a neural network parametrized by $\boldsymbol{\theta} $ can be trained to generate a vector field prediction, denoted as $\bar{v}_t(\mathbf{x}; \boldsymbol{\theta})$, using a simple flow matching objective $\mathcal{L}_{FM}(\boldsymbol{\theta})$:
\begin{equation}
    \mathcal{L}_{FM}(\boldsymbol{\theta}) = \mathbb{E} \| \bar{v}_t\left( \mathbf{x}; \boldsymbol{\theta} \right) - v_t\left( \mathbf{x} \right) \|_2^2
\end{equation}
While $v_t(\mathbf{x})$ and $\mathcal{L}_{FM}(\boldsymbol{\theta})$ are intractable in practice, it has been shown in~\cite{lipman2023flow} that a conditional flow matching objective, denoted as $\mathcal{L}_\text{CFM}(\boldsymbol{\theta})$, can be formulated by conditioning $p_t$ and $v_t$ on real data $\mathbf{x}_1$ and used to train the generative model.
In this work, we follow the optimal transport conditional path proposed in~\cite{lipman2023flow}, which assumes the mean $\mu_t(\mathbf{x}) = t \mathbf{x}_1$ and the standard deviation $\sigma_t(\mathbf{x}) = 1 - \left(1 - \sigma_\text{min} \right) t$. A conditional distribution $p_t \left( \mathbf{x} | \mathbf{x}_1 \right)$ can then be derived as a tractable Gaussian distribution $p_t \left( \mathbf{x} | \mathbf{x}_1 \right) = \mathcal{N} \left( \mathbf{x} | \mu_t(\mathbf{x} \right), \sigma_t^2(\mathbf{x}) \mathbf{I})$ with a corresponding vector field $v_t \left( \mathbf{x} | \mathbf{x}_1 \right) = \frac{\mathbf{x}_1 - \left( 1 - \sigma_\text{min} \right) \mathbf{x}}{1 - \left( 1 - \sigma_\text{min} \right)t}$.
The conditional flow matching objective thus becomes:
\begin{multline}
    \mathcal{L}_{CFM}(\boldsymbol{\theta}) =
    \mathbb{E}\| \bar{v}_t \left( \psi_t(\mathbf{x}_0) ; \boldsymbol{\theta} \right) - \left( \mathbf{x}_1 - \left(1 - \sigma_\text{min} \right) \mathbf{x}_0 \right) \|_2^2
\end{multline}
where $\psi_t(\mathbf{x}_0) = \sigma_t(\mathbf{x}_1)\mathbf{x}_0 + \mu_t(\mathbf{x}_1)$, $t$ is sampled from uniform distribution  $\mathcal{U} \left( 0, 1 \right)$, $\mathbf{x}_0$ is sampled from normal distribution  $\mathcal{N}(\mathbf{0}, \mathbf{I})$, and $\mathbf{x}_1$ is sampled from the target distribution $q(\mathbf{x})$.

A pretraining framework SpeechFlow~\cite{liu2024generative} is proposed for speech generative models using flow matching.
It aims to model the clean speech distribution \( 
q \left( \mathbf{x} \right) \), where the acoustic features \( \mathbf{x} \in \mathbb{R}^d \) represent \( d_\text{mel} \)-dimensional \mbox{mel-spectrograms} of \( L = d / d_\text{mel} \) frames.
A neural vector field estimator parameterized by $\boldsymbol{\theta}$ is trained with a partially masked input \( \mathbf{x}_{\text{mask}} \) as a conditional input when predicting the vector field $\bar{v}_t \left( \psi_t \left( \mathbf{x}_0 \right) | \mathbf{x}_{\text{mask}} ; \boldsymbol{\theta} \right)$ to learn the distribution of clean speech \mbox{mel-spectrograms}.
The pretrained model was then finetuned in speech denoising, speech separation, and text-to-speech synthesis tasks, where \( \mathbf{x}_{\text{mask}} \) is replaced by task-specific inputs such as noisy speech or text embedding~\cite{liu2024generative}. 

\subsection{Proposed model}
While SpeechFlow~\cite{liu2024generative} has proven effective in speech generation, it has several drawbacks.
A major issue is that an extra vocoder is needed in the synthesis stage to decode the generated mel-spectrogram into a time-domain signal.
Furthermore, different vocoders are needed for different downstream tasks, as the vocoder could greatly affect evaluation results.
For example, \mbox{HiFi-GAN}~\cite{Kong2020hifigan} was used for speech synthesis, while a pseudo-inverse mel filterbank was applied for speech denoising to combine the recovered linear magnitude spectrogram with the noisy phase.
A trained three-layer ResNet~\cite{He2016deep} was used for both pseudo-inverse mel transform and phase estimation in speech separation.
This issue complicates the synthesis process, introduces performance limitations due to vocoder selection, and makes it difficult to apply the pretrained model across tasks.

In order to simplify the model design and improve performance, our model operates directly on the compressed complex-valued STFT coefficients.
Specifically, in our work the acoustic features \( \mathbf{x} \in \mathbb{R}^d \) with $d = 2 d_\text{STFT}  L$ represent the stacked real and imaginary components of the $d_\text{STFT}$-dimensional vectors over $L$ frames.
Similar to SpeechFlow, the proposed model uses a 24-layer transformer encoder to predict $\bar{v}_t$.
The time embedding, obtained via sinusoidal position encoding of the time step $t$, was originally appended to the input along the time axis~\cite{liu2024generative}.
However, this can prevent far-end frames from utilizing time information due to the time-distance penalty caused by ALiBi self-attention bias~\cite{Press2022alibi}.
Our model adopts the adaptive normalization layer as in diffusion transformers~\cite{Peebles2023scalable}, adding 100M trainable parameters and resulting in a total model size of 430M.
Larger models with approximately 860M and 1.1B parameters were also pretrained and finetuned, but yielded no significant improvements in most tasks, with marginal gains observed in the TSE task.
At inference time, we start from a sample $\mathbf{x}_0$, solve the ODE as in~\eqref{eq:ode} with vector field prediction $\bar{v}_t \left( . | \mathbf{x}_\text{cond} ; \boldsymbol{\theta} \right)$ to obtain an estimate $\hat{\mathbf{x}}_1$, and obtain the time-domain estimate by applying the inverse STFT.
\section{Experimental setup}
\label{sec:experimental_setup}
\subsection{Pretraining dataset and setup}
We use the \mbox{Libri-Light} dataset~\cite{Kahn2020librilight} with approximately 60k hours of English speech at 16kHz sample rate for pretraining.
The model is trained on 32 NVIDIA H100 GPUs with a batch size of 131 seconds per GPU for 600k steps.
We use the Adam optimizer, warming up the learning rate linearly to $5 \cdot 10^{-5}$ for the first 5k steps, and then decaying to $10^{-5}$ using a cosine annealing scheduler. 
The STFT coefficients is computed using the STFT window size of 510 samples, hop size of 128 samples, and the compression parameters $a = 0.5$ and $b = 0.33$~\cite{lemercier2023storm}. 
The minimum standard deviation $\sigma_\text{min}$ is set to $10^{-4}$.
The masked condition $x_\text{mask}$ is obtained by randomly selecting $70\%$ of frames across time to be
masked to zero, with a minimum masking span length of ten frames.
Note that we experimented with alternative masking strategies, such as masking time-frequency patches or using random masking values instead of zeros.
However, the considered masking strategies showed similar performance after finetuning, so alternative results are not reported. 
During pretraining,  we drop the conditional inputs with a $10\%$ probability to perform an unconditioned flow matching generative process.
\subsection{Finetuning tasks and setup}
Four tasks are considered in our experiments: speech denoising, bandwidth extension, codec artifact removal, and target speaker extraction. All tasks were finetuned and evaluated on 16kHz speech signals. 
The unprocessed signal is used as the conditional input $\mathbf{x}_\text{cond}$, unless noted otherwise.

\noindent \textbf{Speech Denoising:} We used the \mbox{VoiceBank-DEMAND} (\mbox{VB-DMD})~\cite{ValentiniBotinhao2017} and \mbox{WSJ0-CHIME3}~\cite{lemercier2023storm} datasets. For \mbox{VB-DMD} dataset, we followed the common train, validation and test partitions proposed in \cite{Lu2022conditional}, resulting in 10,802 training utterances, 770 validation utterances and 824 evaluation utterances.
The WSJ0-CHiME3 dataset was prepared similarly to~\cite{lemercier2023storm} with 12,776 utterances ($\sim$25h) for training, 1,206 utterances ($\sim$ 2h) for validation and 651 utterances ($\sim$1.5h) for testing.

\noindent \textbf{Bandwidth Extension:} We prepared the \mbox{WSJ0-BWE} dataset following the process in~\cite{lemercier2023analysing}, where the input speech is decimated by a down-scaling factor sampled uniformly in \{2, 4, 8\} and then resampled to 16kHz.
Similar to the \mbox{WSJ0-CHiME3} dataset, it consists of 12,776 utterances for training, 1,206 utterances for validation, and 651 utterances for testing.

\noindent \textbf{Codec Artifact Removal:} The goal in this task is to restore high-quality speech from low-bitrate audio codec outputs.
We used the publicly available 16kHz pretrained Descript Audio Codec (DAC) model~\cite{kumar2023highfidelity} with a maximum of 12 codebooks and bitrate of 6kbps.
Low-bitrate input signals were generated using the first four codebooks, resulting in a 2kbps bitrate. 
The WSJ0 dataset was encoded and decoded using the four-codebook DAC, producing degraded speech with significant coding artifacts. 
The pretrained model was finetuned to restore the original speech from this degraded speech as input.
For reference, we also generated the dataset using either 8 or all 12 codebooks for comparison.

\noindent \textbf{Target Speaker Extraction:} This task involves extracting the target speaker’s speech while removing interfering speakers. 
We uses the Libri2Mix dataset~\cite{cosentino2020librimix}, consisting of simulated mixtures of two speakers in the min mode at a 16kHz sample rate.
Following the data preparation in \mbox{TD-SpeakerBeam}~\cite{delcroix2020improving}, the dataset was partitioned into three subsets: \mbox{train-100} set (27,800 utterances), validation set (6,000 utterances), and test set (6,000 utterances). 
Note that in the TSE task, in addition to the mixture speech as corrupted speech observation, a reference speech signal is also needed as a target speaker prompt.
In order to provide the target speaker information, we trim the first three seconds of the target reference speech, prepend it to the input mixture signal to form the conditional input $\mathbf{x}_\text{cond}$.
Finally, we remove the first three seconds of the generated output, keeping only the portion corresponding to the input mixture.

All finetuning processes are performed on a single NVIDIA H100 GPU for 160 epochs with a batch size of 50 seconds.
The learning rate peaks at $2 \cdot 10^{-5}$ after 5k updates, then decay to 0 with a cosine annealing scheduler.
For the control group without pretraining (training from scratch), the learning rate peaks at $10^{-4}$ after 5k updates and decay to 0. Note that in the finetuning stage we always feed the conditional input to the model, as classifier-free guidance~\cite{ho2021classifierfree} did not improve speech quality in our experiments.
For the ODE solver, we use the Euler method to compute $\phi_1(x_0)$ from $\phi_0(x_0)$ by approximating the integration from $t = 0$ to $t = 1$ with a step size of $\Delta t = 0.2$, resulting in only five neural model evaluations to generate the output signal.

\subsection{Evaluation metrics}
Our evaluation metrics include perceptual evaluation of speech quality (PESQ)~\cite{rix2001pesq}, short-term objective intelligibility (STOI)~\cite{Jensen2016stoi} and its extended version (eSTOI), and scale-invariant signal-to-distortion ratio (SI-SDR)~\cite{Roux2019sisdr}.
However, as noted in~\cite{lemercier2023analysing}, we found that the aforementioned instrumental metrics are sometimes inconsistent with human listening evaluations particularly in the bandwidth extension and codec artifact removal tasks.
To better assess speech quality, we complemented the metrics with \mbox{SQUIM-MOS}~\cite{kumar2023torchaudio} and \mbox{WV-MOS}~\cite{Andreev2023hifi}.
In the TSE task, we also included the failure rate (FR)~\cite{delcroix2022listen}, which measures the proportion of test samples with an SI-SDR improvement below 1~dB and indicates how often the TSE system fails to extract the correct speaker.
\section{Results}
\label{sec:results}

In this section we report the results of different tasks in terms of the instrumental metrics.
Audio examples can be found on the demo page\footnote{\url{https://kuray107.github.io/ssl_gen25-examples/index.html}}, where we randomly selected five examples from the test set for each task evaluated in the paper.
We encourage the reader to listen to the generated audios to validate the results presented in this section.
\begin{table}[t]
\caption{Speech denoising performance on Voicebank-DEMAND dataset. $^\dagger$denotes results taken from corresponding prior works.}
\vspace{-2.5em}
\label{table:vbdmd}
\begin{center}
  \setlength\tabcolsep{2.0pt} 
  \resizebox{\columnwidth}{!}{%
    \begin{tabular}{ccccc}
      \toprule
      Signal      & WV-MOS~$\uparrow$ & PESQ~$\uparrow$ & ESTOI~$\uparrow$ & SI-SDR/dB~$\uparrow$ \\ 
      \midrule
      Clean                     & 4.50 & --    & --    & --     \\ 
      Unprocessed               & 2.98 & 1.97  & 0.79  & 8.4    \\ 
      \midrule
      SGMSE+$^\dagger$~\cite{richter2023sgmse}                     & 4.24 & 2.93  & 0.87  & 17.3   \\ 
      StoRM$^\dagger$~\cite{lemercier2023analysing}                & 4.30 & 2.93  & \textbf{0.88}  & 18.8   \\
      VPIDM$^\dagger$~\cite{guo2024variance}                       & --   & 3.16  & 0.87  & --     \\
      SpeechFlow$^\dagger$~\cite{liu2024generative}                & --   & 3.13  & 0.87  & --     \\ 
      SpeechFlow (w/o pretrain)$^\dagger$~\cite{liu2024generative} & --   & 2.93  & 0.85  & --     \\ 
      \midrule
      Proposed                  & \textbf{4.41} & \textbf{3.27} & \textbf{0.88} & 19.1 \\ 
      Proposed (w/o pretrain)   & 4.27 & 3.05 & 0.87 & 18.7 \\ 
      \bottomrule
    \end{tabular}
  }
\end{center}
\end{table}

\begin{table}[t]
\caption{Speech denoising performance on \mbox{WSJ0-CHiME3}.}
\vspace{-2.5em}
\label{table:wsj0_chime}
\begin{center}
  \setlength\tabcolsep{2.0pt} 
  \resizebox{\columnwidth}{!}{%
    \begin{tabular}{ccccc}
      \toprule
      Signal      & WV-MOS~$\uparrow$ & PESQ~$\uparrow$ & ESTOI~$\uparrow$ & SI-SDR/dB~$\uparrow$ \\ 
      \midrule
      Clean       & 4.16  & -- & -- & --  \\ 
      Unprocessed & 1.42* & 1.35 & 0.63 & 4.0  \\ 
      \midrule
      SGMSE+~\cite{richter2023sgmse}      & 3.64 & 2.28 & 0.85 & 13.1  \\ 
      StoRM~\cite{lemercier2023analysing} & 3.72 & 2.53 & 0.87 & 14.8 \\ 
      \midrule
      Proposed    & \textbf{4.26} & \textbf{2.85} & \textbf{0.92} & \textbf{16.2} \\ 
      Proposed (w/o pretrain) & 4.15 & 2.59 & 0.90 & 15.6 \\ 
      \bottomrule
    \end{tabular}
  }
\end{center}
\footnotesize *Some unprocessed signals resulted in negative WV-MOS scores. Since MOS cannot be below 1.0, we applied a lower bound of 1.0 to obtain the reported average score, virtually matching 1.43 reported in~\cite{lemercier2023analysing}.
\end{table}

\noindent \textbf{Speech Denoising:} 
We first assess the proposed model's performance on the speech denoising task and compare it with other generative-based models, including SpeechFlow~\cite{liu2024generative}, SGMSE+~\cite{richter2023sgmse}, StoRM~\cite{lemercier2023storm}, and VPIDM~\cite{guo2024variance}.
Table~\ref{table:vbdmd} shows the test performance of the models on the VB-DMD dataset. 
The finetuned model outperformed the existing generative-based models in all evaluation metrics, particularly in PESQ and WV-MOS, demonstrating superior speech quality.
Additionally, the proposed model surpasses SpeechFlow in both pretraining and non-pretraining configurations.
This verifies our hypothesis that working in the \mbox{mel-spectrogram} domain can limit model performance. 
Similar observations can be made in Table~\ref{table:wsj0_chime}, which reports the test performance on the WSJ0-CHiME3 dataset. 
Again, the finetuned model outperforms existing diffusion-based models across all metrics by a large margin.

\begin{table}[t]
\caption{Bandwidth Extension performance on WSJ0-BWE dataset. $^\dagger$denotes results taken from~\cite{lemercier2023analysing}.}
\vspace{-2.5em}
\label{table:bwe}
\begin{center}
  \setlength\tabcolsep{2.0pt} 
  \resizebox{\columnwidth}{!}{%
    \begin{tabular}{ccccc}
      \toprule
      Signal                  & WV-MOS~$\uparrow$ & SQUIM-MOS~$\uparrow$ & ESTOI~$\uparrow$ & SI-SDR/dB~$\uparrow$ \\ 
      \midrule
      Clean                   & 4.16  & 4.46  & --   & --   \\ 
      Unprocessed             & 2.45  & 2.94  & 0.72 & -3.8 \\ 
      \midrule
      NCSN++$^\dagger$~\cite{lemercier2023analysing} & 2.25  & --    & 0.73 & --   \\ 
      SGMSE+$^\dagger$~\cite{richter2023sgmse}       & 3.43  & --    & 0.83 & --   \\ 
      \midrule
      Proposed                & \textbf{3.93} & \textbf{4.41} & \textbf{0.92} & \textbf{8.5} \\ 
      Proposed (w/o pretrain) & 3.58  & 4.35 & 0.89 & 6.2  \\ 
      \bottomrule
    \end{tabular}
  }
\end{center}
\end{table}

\noindent \textbf{Bandwidth Extension:}
Table~\ref{table:bwe} compares our model's bandwidth extension performance with other generative models. Our model consistently demonstrate great ability in restoring high-frequency information, outperforming the baseline models in both eSTOI and WV-MOS evaluations. The SQUIM-MOS score further confirms its ability to generate speech that closely matches the ground truth. Note that while the proposed model without pretraining already outperforms the baselines, pretraining significantly enhances performance across all metrics, particularly in SI-SDR.

\begin{table}[t]
\caption{Codec artifact removal performance on WSJ0-DAC dataset.}
\vspace{-1.5em}
\label{table:dac}
\begin{center}
  \setlength\tabcolsep{2.0pt} 
  \resizebox{\columnwidth}{!}{%
    \begin{tabular}{ccccc}
      \toprule
      Signal & WV-MOS~$\uparrow$ & SQUIM-MOS~$\uparrow$ & ESTOI~$\uparrow$ & SI-SDR/dB~$\uparrow$ \\ 
      \midrule
      Clean                   & 4.16  & 4.46  & --    & --     \\ 
      4-codebook DAC (Input)    & 2.80  & 3.47  & 0.70  & -4.6   \\ 
      \midrule
      8-codebook DAC          & 3.75  & 4.12  & 0.87  & 4.1    \\
      12-codebook DAC         & 4.00  & 4.46  & 0.95  & 8.2    \\
      \midrule
      Proposed                & 4.02  & 4.44  & 0.86  & 3.5    \\
      Proposed (w/o pretrain) & 4.10  & 4.43  & 0.84  & 2.8    \\ 
      \bottomrule
    \end{tabular}
  }
\end{center}
\end{table}

\noindent \textbf{Codec Artifact Removal:} 
Table~\ref{table:dac} presents the results of the codec artifact removal task, comparing the input speech encoded and decoded with four codebooks, the processed speech from our model, and speech encoded and decoded with 8 and 12 codebooks.
The DAC model generates high-quality speech that closely resembles the original with all 12 codebooks.
However, reducing the number of codebooks to eight or four significantly degrades the generated speech quality.
The proposed finetuned model effectively mitigates the codec artifacts and restores the speech quality.
In eSTOI and SI-SDR evaluations, the generated speech scores similarly to the 8-codebook DAC.
More importantly, in WV-MOS and SQUIM-MOS evaluations, which we found align better with human judgment, our model generates speech with scores comparable to those of the 12-codebook DAC.
This suggests that low-bit-rate audio codecs can be employed for speech transmission or generation, with our model serving as a post-processor to recover quality.
Additionally, \mbox{Speech-LLM} and codec-based TTS models can benefit from this approach, as it enables the generation of fewer codec tokens and speeding up training and testing with our model as an add-on to maintain speech quality.

\begin{table}[t]
\caption{Target Speaker Extraction performance on Libri2mix train-100 (16k, min) dataset. $^\dagger$denotes results taken from~\cite{peng2024target}.}
\vspace{-2.5em}
\label{table:libri2mix}
\begin{center}
  \setlength\tabcolsep{3.0pt} 
  \resizebox{\columnwidth}{!}{%
    \begin{tabular}{ccccc}
      \toprule
      Signal                  & PESQ~$\uparrow$ & STOI~$\uparrow$ & SI-SDRi~$\uparrow$ & FR/\%~$\downarrow$ \\ 
      \midrule
      Unprocessed             & 1.15            & 0.71            & 0.00              & -- \\ 
      \midrule
      TD-SpeakerBeam
      $^\dagger$~\cite{delcroix2020improving}          & 2.12            & 0.90            & 13.03             & 4.8 \\ 
      SpEx+$^\dagger$~\cite{ge2020spexcompletetimedomain}                   & 2.93            & --              & 13.41             & -- \\ 
      sPDCCN$^\dagger$~\cite{han2022dpccndenselyconnectedpyramidcomplex}                   & --              & --              & 11.61             & -- \\ 
      SSL-MHFA$^\dagger$~\cite{peng2024target}                & 2.45            & 0.93            & 14.65             & 3.0 \\ 
      SSL-MHFA (w/o finetune)$^\dagger$~\cite{peng2024target} & 2.38            & 0.91            & 14.01             & 3.9 \\ 
      \midrule
      Proposed                & \textbf{2.99}   & \textbf{0.94}   & \textbf{16.00}             & \textbf{2.4} \\ 
      Proposed (w/o pretrain) & 2.52            & 0.92            & 13.46             & 4.6 \\
      \bottomrule
    \end{tabular}
  }
\end{center}
\end{table}

\noindent \textbf{Target Speaker Extraction:} 
Finally, we evaluate our model's effectiveness in the TSE task. 
Our baseline models include TD-SpeakerBeam~\cite{delcroix2020improving}, SpEx+~\cite{ge2020spexcompletetimedomain}, sPDCCN~\cite{han2022dpccndenselyconnectedpyramidcomplex}, and SSL-MHFA~\cite{peng2024target}.
Note that SSL-MHFA finetunes a pretrained WavLM~\cite{chen2022wavlm} as an acoustic feature extractor, making it an ideal comparison to our model. 
As shown in Table~\ref{table:libri2mix}, the proposed model outperforms all baselines in all evaluation metrics by a large margin, once again demonstrating its superior speech restoration ability. 
Furthermore, all the above baseline models require an extra speaker encoder to extract speaker embeddings from the reference speech, whereas our model directly processes the raw reference and mixture speech together without relying on a single speaker embedding. 
As a result, our proposed model achieves the lowest FR with a 20\% relative improvement compared to the current state-of-the-art system, SSL-MHFA~\cite{peng2024target}.
\section{Conclusion}
In this paper, we present a foundational generative model for high-quality speech restoration based on flow matching. 
The proposed model operates on complex-valued STFT coefficients, significantly simplifying the synthesis process compared to prior work.
We finetuned the proposed model on four different speech restoration tasks to demonstrate its effectiveness.
In speech denoising and bandwidth extension tasks, the proposed model outperforms SpeechFlow and existing generative baseline models in all evaluation metrics.
In the codec artifact removal task, our model can recover speech quality close to the original speech from a low-bit rate codec output.
In the target speaker extraction task, our model outperforms existing TSE systems and achieves a new state-of-the-art performance. 
With our implementation and pretrained checkpoints release in the NVIDIA NeMo Framework~\cite{nemo_toolkit}, we hope this work can serve as a general foundational model for future speech restoration and generation tasks. 

%
\cleardoublepage
%
\bibliographystyle{IEEEtran}
\bibliography{references/refs}

\end{document}